\documentclass[twocolumn,aps,prd,epsf]{revtex4}
\usepackage{epsfig}
\usepackage{amsmath}

\begin{document}

\title{The Charmonium (Bottomonium) binding ate finite T}

\author{P. Bicudo
%\email{}
}
\author{M. Cardoso
%\email{}
}
\author{P. Santos
%\email{}
}
\author{J. Seixas
%\email{}
}
\affiliation{CFTP, Dep. F\'{\i}sica, Instituto Superior T\'ecnico,
Av. Rovisco Pais, 1049-001 Lisboa, Portugal}

\begin{abstract}
The charmonium (bottomonim) binding at finite temperature is
studied with static potentials extracted from
the lattice QCD data of Kaczmarek {\em et al}.  
The bottomonium spectrum is also 
studied. This is relevant for Hard Probes
in Heavy Ion Collisions.
\end{abstract}
\maketitle

%%%%%%%%%%%%%%%%%%%%%%%%%%%%%%%%%%%%%%%%%%%%%%%%%%%%%%%%%%%%%%%%%%%%%%%
%%                                                                   
%%
%%                                   SSS                             
%%
%%                                  S   S                            
%%
%%                                   S                               
%%
%%                                    S                              
%%
%%                                     S                             
%%
%%                                  S   S                            
%%
%%                                   SSS                             
%%
%%                                                                   
%%
%%%%%%%%%%%%%%%%%%%%%%%%%%%%%%%%%%%%%%%%%%%%%%%%%%%%%%%%%%%%%%%%%%%%%%%

\section{Introduction}

This talk is motivated by the seminal hard $c \bar c$ probe paper
of Matsui and Satz
\cite{Matsui:1986dk}
by the lattice QCD data on finite temperature static potentials 
\cite{Doring:2007uh,Hubner:2007qh,Kaczmarek:2005ui,Kaczmarek:2005gi,Kaczmarek:2005zp,Karsch:2004ik,Petreczky:2004pz,Nakamura:2004wra}
and binding of charmonia
\cite{Asakawa:2003re,Datta:2003ww}, 
and by the theoretical studies
\cite{Wong:2004zr,Shuryak:2004tx,Mocsy:2004bv,Digal:2001iu}
on finite temperature lattice potentials.

With a finite $T$ quark potential, and modern quark model techniques one might,
approximately,
\\
- study chiral symmetry breaking, quark mass generation, a finite $T$
notice that if one maintains a confining potential, chiral symmetry is always broken, whatever the $T$,
\\
- compute the spectrum of any hadron at finite $T$
not only the $J/\psi$  but also light mesons, baryons, etc
\\
- compute the interaction of any hadron-hadron  at finite $T$
using cluster methods like the Resonating Group Method.

\begin{figure}[b]
%\begin{picture}(230,145)(0,0)
%\put(0,0){
\includegraphics[width=0.99 \columnwidth]{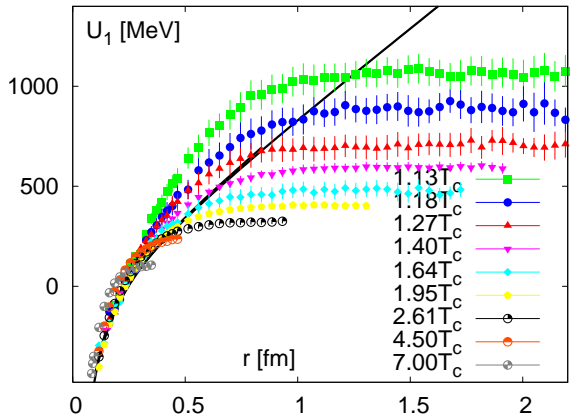}
%}
%\end{picture}
\caption{\label{U1Kacz}Lattice QCD data for the internal energy $U_1$, thanks to Olaf Kaczmarek et al. 
The solid line represents the $T=0$ potential .}
\end{figure}

Here just study the charmonium and bottomonium as prototypes to study  finite T quark potentials.

The charmonium is a good starting point because 
\begin{eqnarray}
m_c  &>>& \Lambda_{QCD} \ , 
\nonumber \\
m_c  &>>& T_c  \ .
\end{eqnarray}  
Thus it is reasonable to neglect in the bound state equation,
spontaneous chiral symmetry breaking,
relativistic effects,
coupled channels,
and  temperature effects (other than the potential dependence on the temperature) .
We can simply solve the Schrödinger equation with static lattice QCD finite $T$ potentials

\begin{figure}[t]
%\begin{picture}(230,145)(0,0)
%\put(0,0){
\includegraphics[width=0.99 \columnwidth]{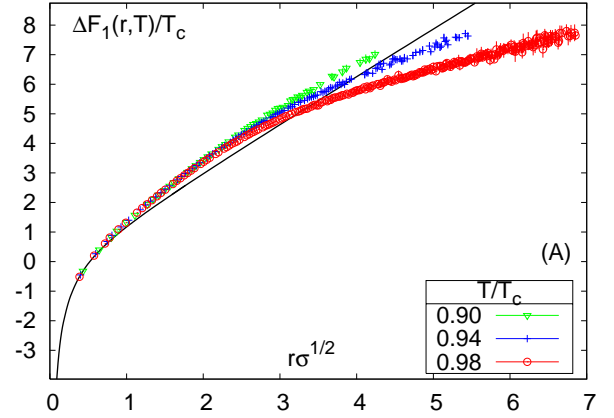}
\\
\vspace{1.cm}
\includegraphics[width=0.99 \columnwidth]{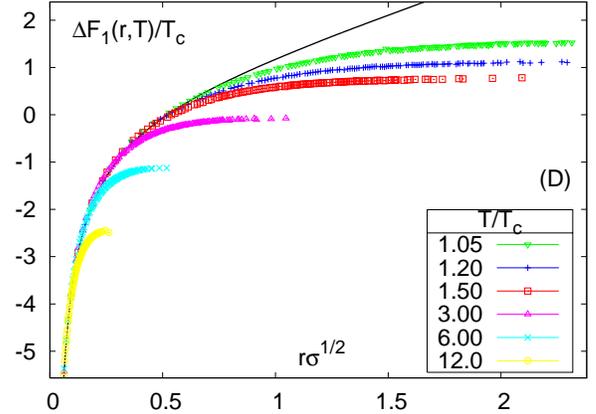}
%}
%\end{picture}
\caption{\label{F1Kacz}Lattice QCD data for the free energy $F_1$, thanks to Olaf Kaczmarek et al.
 The solid line represents the $T=0$ potential .}
\end{figure}

\begin{figure*}[t]
%\begin{picture}(230,235)(0,0)
\includegraphics[width=0.9\columnwidth]{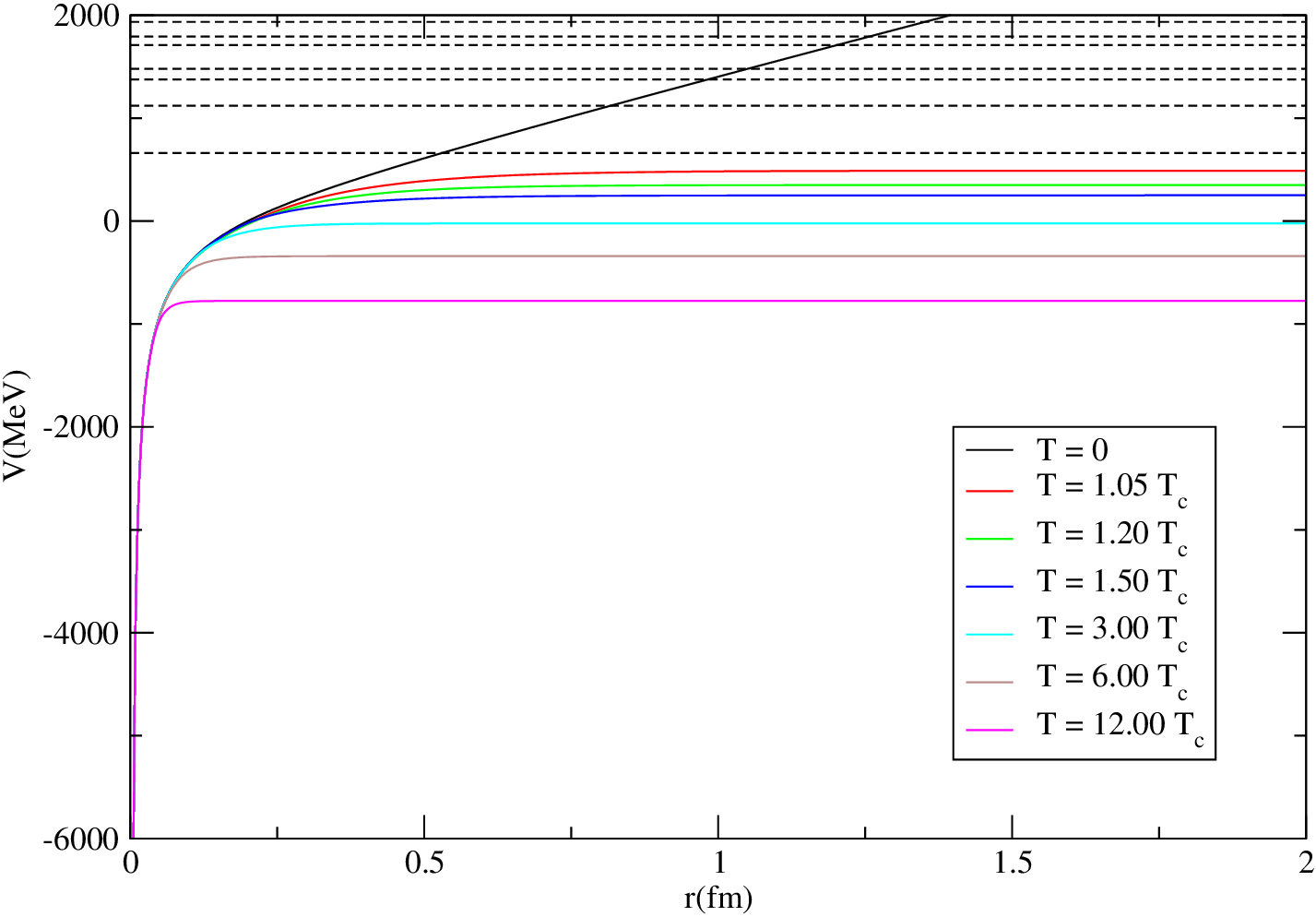}
\hspace{1.cm}
\includegraphics[width=0.8\columnwidth]{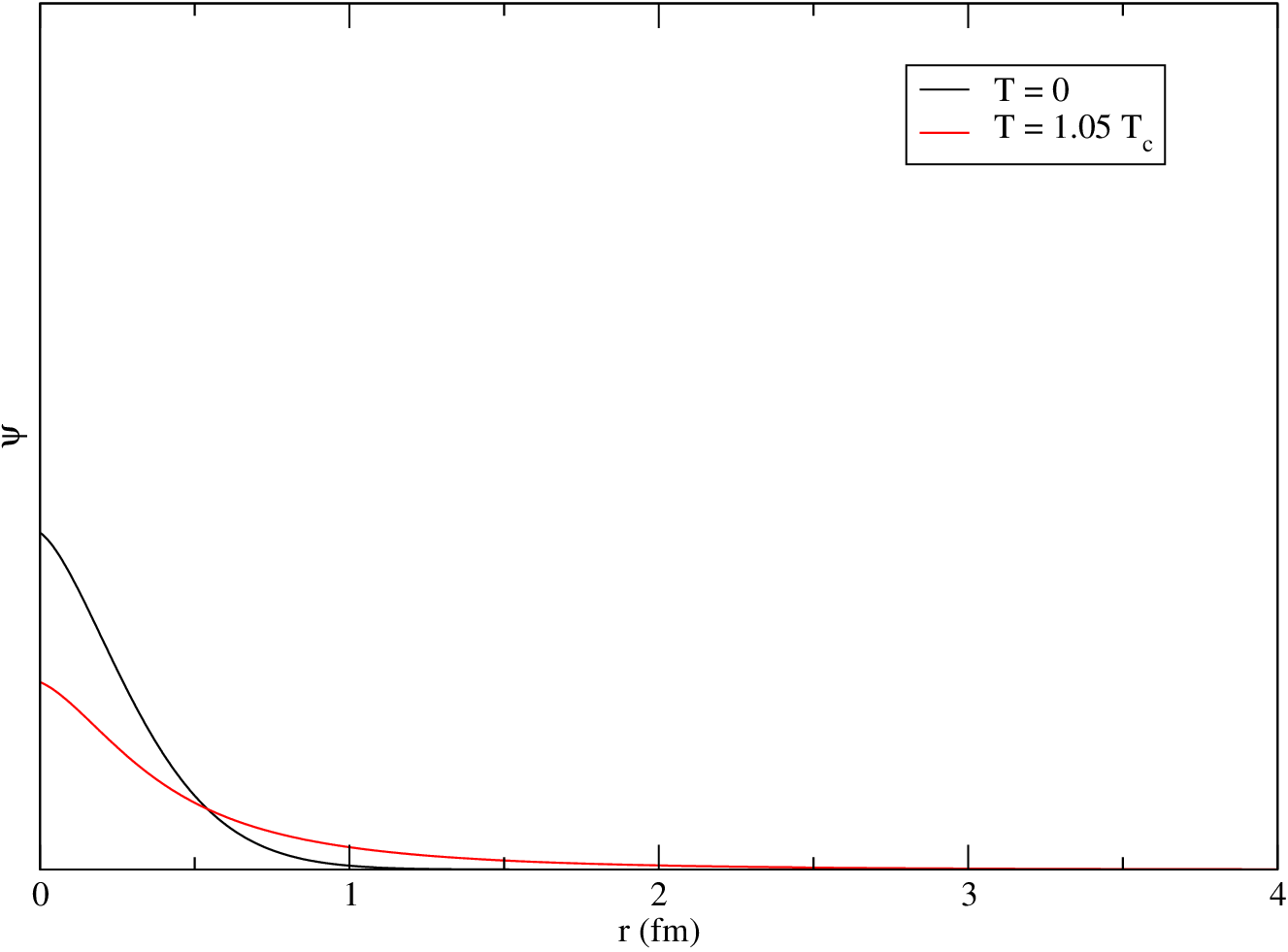}
\\  
\vspace{1.cm}
\includegraphics[width=0.9\columnwidth]{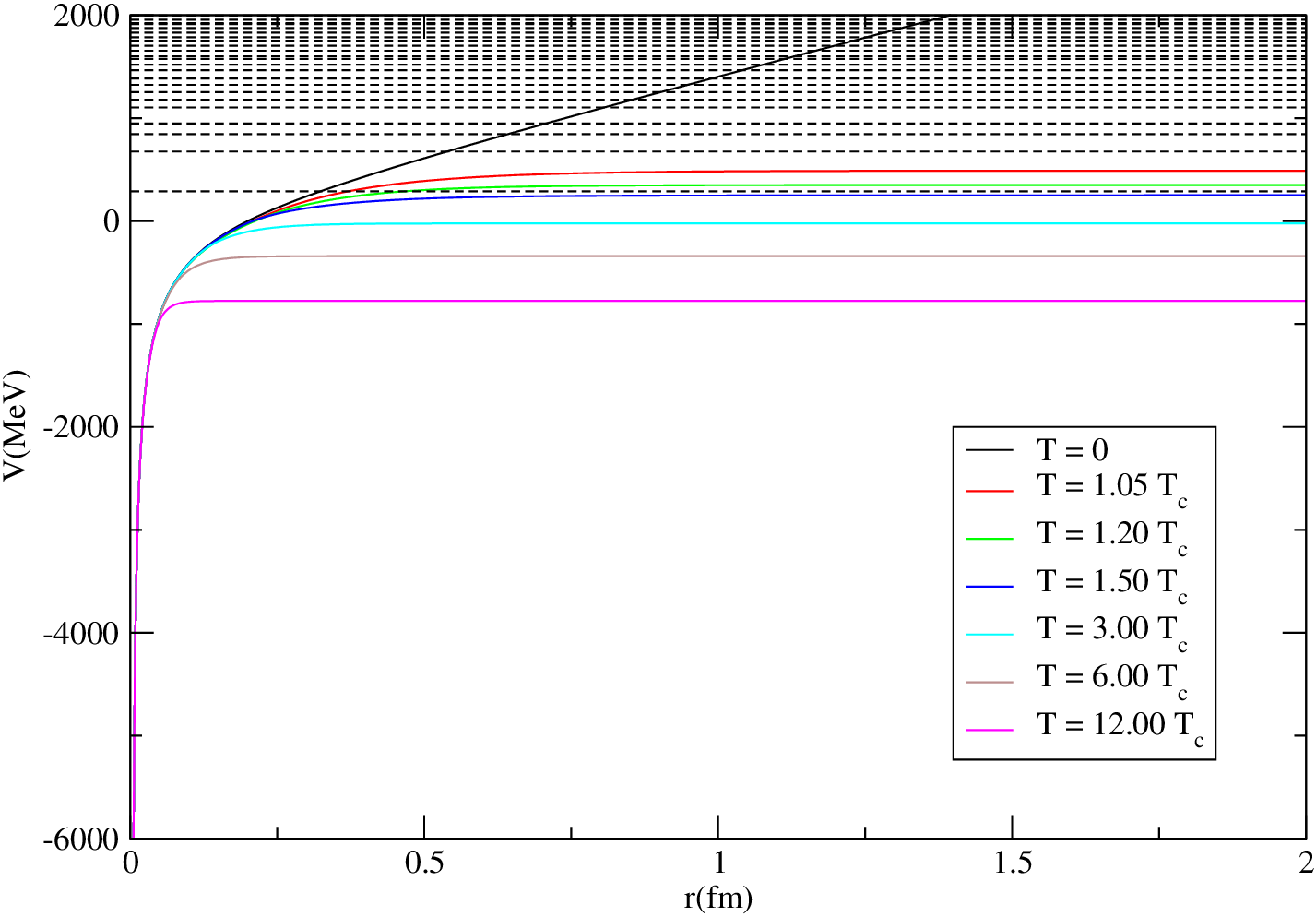}
\hspace{1.cm}
\includegraphics[width=0.8\columnwidth]{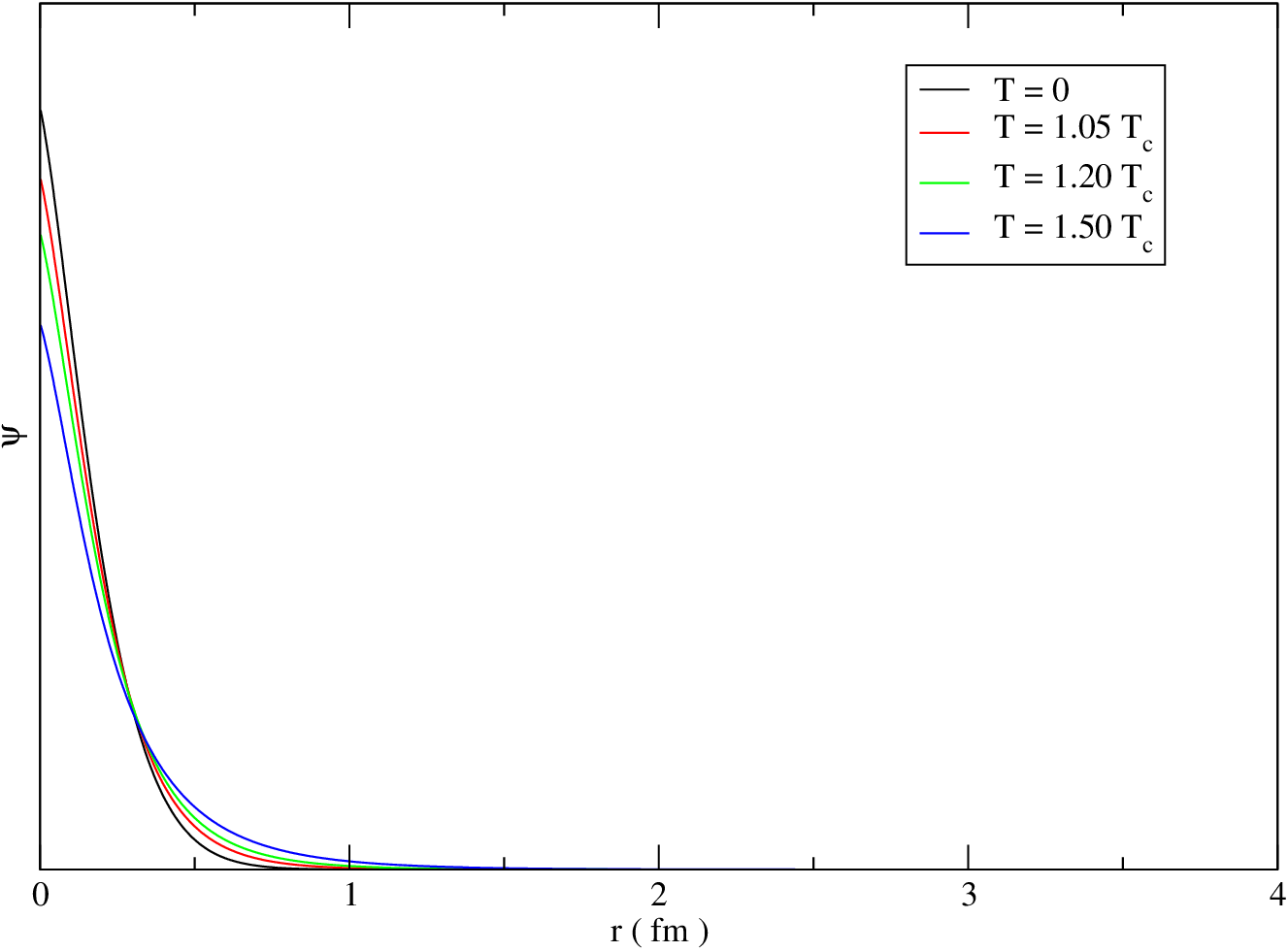}
%\end{picture}
\caption{\label{freeenvelope}Boundstates with the Free Energy potentials. 
In the top left we show the energy spectrum of charmonium in the enveloping potential, and in the top right we 
show the wavefunctions for the ground state. In the bottom left we show the energy spectrum of 
bottomonium in the enveloping potential, and in the bottom right we show the wavefunctions for the ground state.}
\end{figure*}

\begin{table}[b]
	\begin{tabular}{ccccc}
	\hline
	$c$ & $\alpha$ & $\alpha'$ & $\Lambda$ & $\sigma$ \\
	\hline
   -0.0270 & 0.2857 & 0.1333 & 10.00 & 1.1621 \\
	\hline 
	$T / T_c$ & $d$ & $A$ & $\lambda$ & $w$  \\
\hline
	1.05 & 0.98 & 0.357 & 2.1 &  0.10   \\
	1.20 & 0.70 & 0.340 & 2.7 &  0.05   \\
	1.50 & 0.50 & 0.537 & 2.7 &  0.10    \\
	3.00 & -0.045 & 0.301 & 6.0  & 0.05    \\
	6.00 & -0.68 & 0.239 & 11.0  & 0.10    \\
	12.0 & -1.55 & 0.116 & 25.0  & 0.10    \\
		\end{tabular}
\caption{ Fitting parameters of the  free energy  ${F_1}_T(r)$ .
\label{free energy parameters}}
\end{table}

\begin{table}[b]
	\begin{tabular}{ccccccccc}
	\hline
	& $c$ & $\alpha$ & $\alpha'$ & $\Lambda$ & $\sigma$ & \\
	\hline
& -0.2117 & 0.3490 & 0.1733 & 7.793 & 4.475  & \\
	\hline
 $T / T_c$ & $d$ & $A$ & $\lambda$ & $A'$ & $\lambda'$  & $w$ \\
\hline
 1.13 & 3.94 & 6.99 & 2.64 &0&0&  0.336  \\
  1.18 & 2.35 & 4.58 & 2.27  &0&0&  0.847 \\
  1.27 & 2.64 & 5.38 & 2.87  &0&0&  0.461  \\
   1.40 & 2.23 & 5.57 & 2.91   &0&0&  0  \\
  1.64 & 1.79 & 5.60 & 3.65   &0&0& 0 \\
 1.95 & 1.51 & 5.57 & 4.15   &0&0& 0.117  \\
  2.61 & 1.21 & 6.12 & 5.44  &0&0& 0 \\
   4.50 & -40.7 & 12.6 & 5.44  &1.00 &0.018&  0 \\
   7.00 & -74.0 & 15.9 & 5.44  & 0.49 & 0.021&  0 \\
		\end{tabular}
\caption{ Fitting parameters of the internal energy  ${U_1}_T(r)$ .
\label{internal energy parameters}}
\end{table}

\begin{figure*}[t]
%\begin{picture}(230,235)(0,0)
\includegraphics[width=0.9\columnwidth]{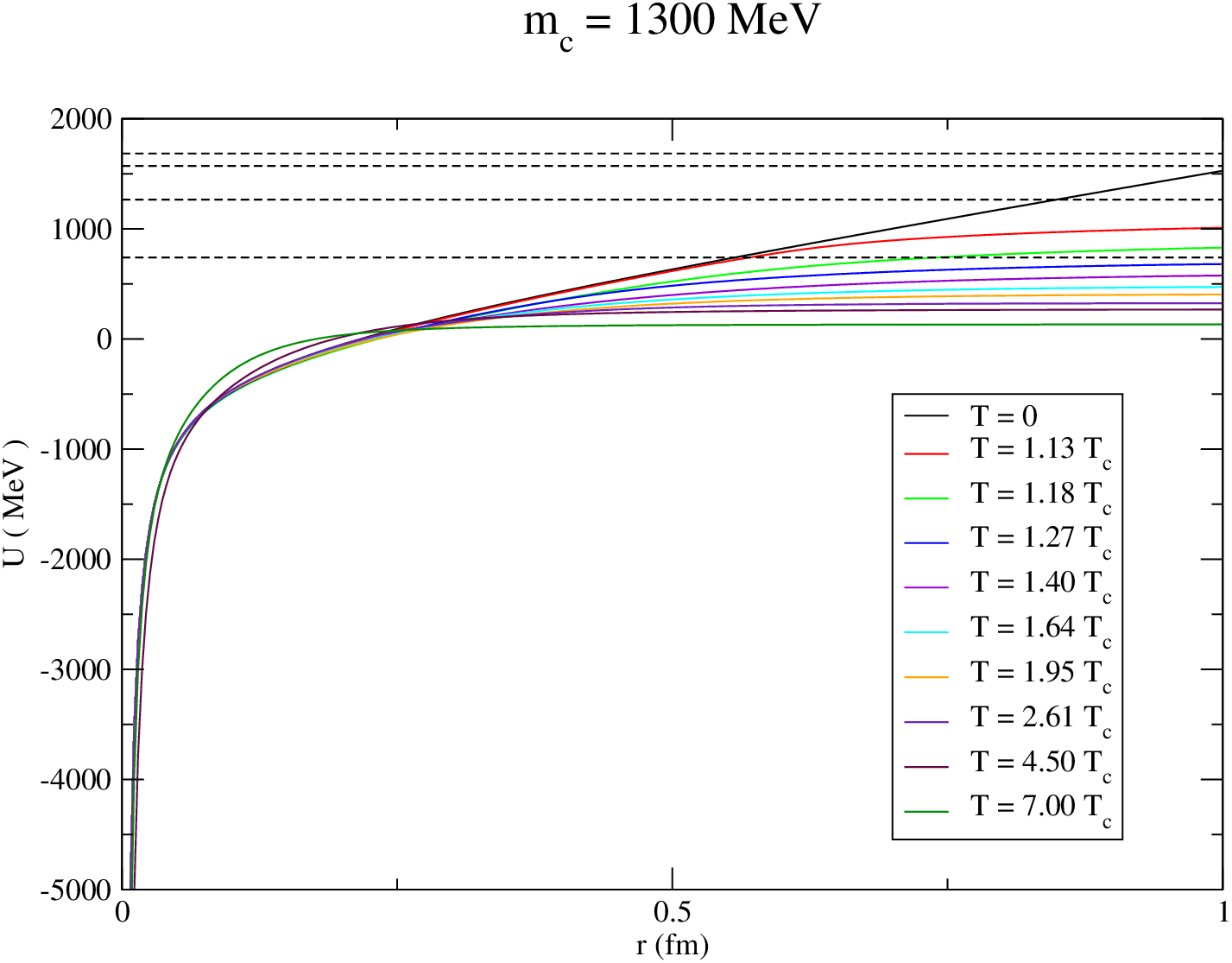}
\hspace{1.cm}
\includegraphics[width=0.8\columnwidth]{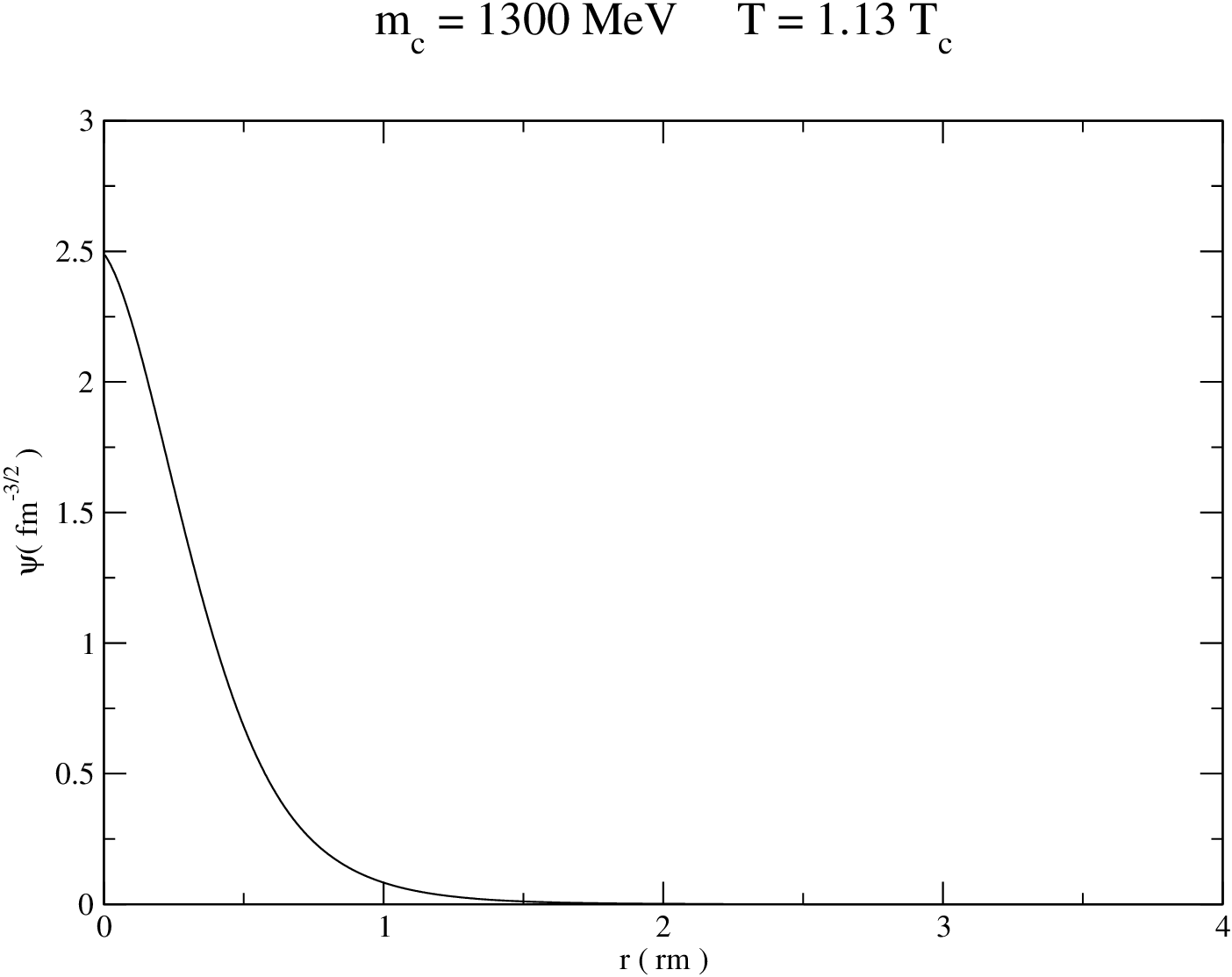}
\\ 
\vspace{1.cm}
\includegraphics[width=0.9\columnwidth]{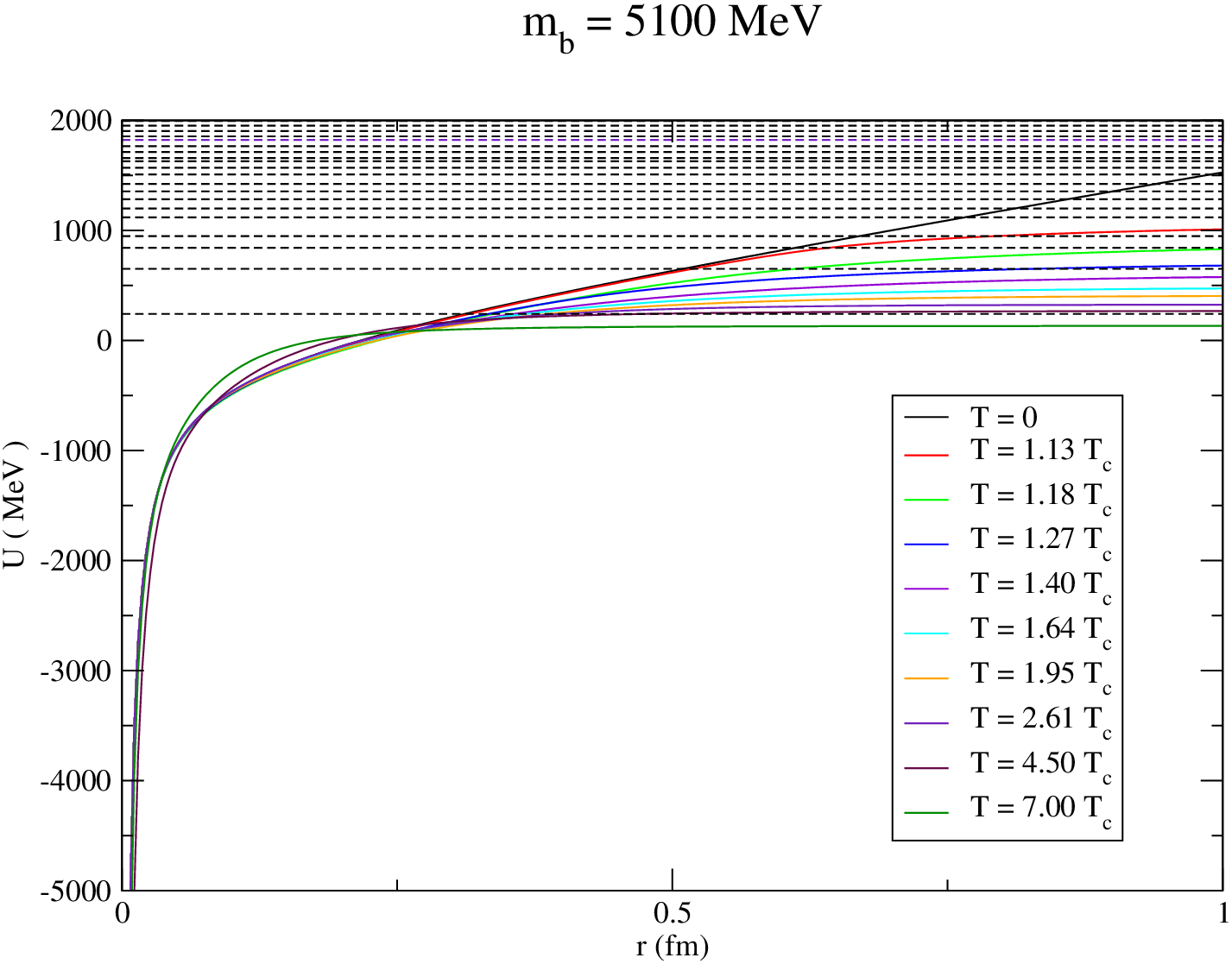}
\hspace{1.cm}
\includegraphics[width=0.8\columnwidth]{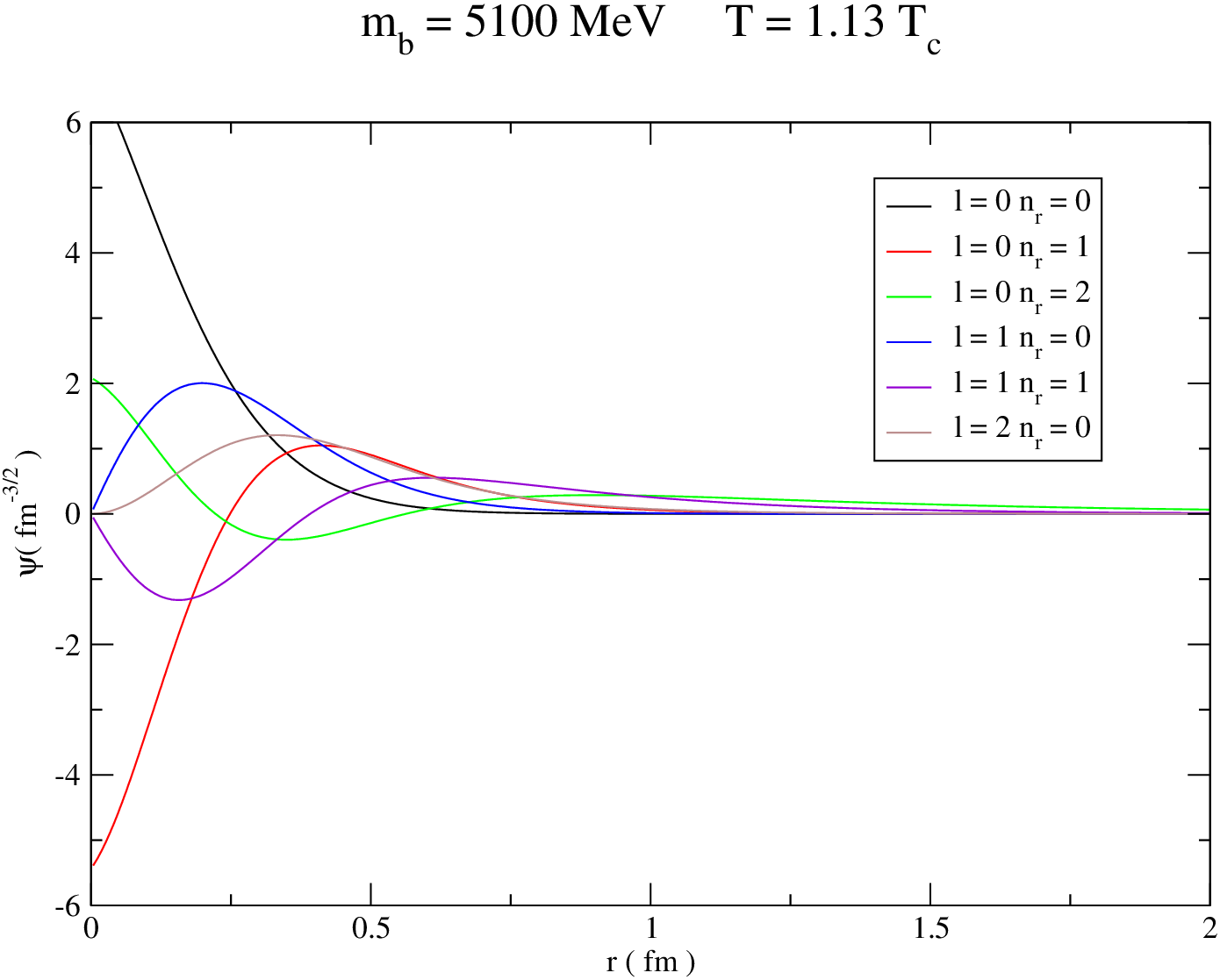}
%\end{picture}
\caption{\label{internalenvelope}Boundstates with the Internal Energy potentials. 
In the top left we show the energy spectrum of charmonium in the enveloping potential, and in the top right we 
show the wavefunctions for the ground state.In the bottom left we show the energy spectrum of 
bottomonium in the enveloping potential, and in the bottom right we show the wavefunctions for the ground state.}
\end{figure*}

\begin{table}[b]
	\begin{tabular}{cccc}
	\hline
	$T/Tc$ & $E_{00} ( MeV )$ & $B(MeV)$ & $\sqrt{\langle r^2 \rangle} (fm)$ \\
	\hline
	0 & 660 & $-\infty$  & 0.390 \\
	1.05 & 465 & $-24.7$ & 0.921 \\
	1.17 & 465 & $-0.593$ & 4.875 \\
	\end{tabular}
	\caption{\label{charm_table} 
Goundstate solutions of the charmonium at different temperatures, for the free energy 
${F_1}_T(r)$ . }
\end{table}

\begin{table}[b]
	\begin{tabular}{cccc}
	\hline
	$T/Tc$ & $E_{00} ( MeV )$ & $B(MeV)$ & $\sqrt{\langle r^2 \rangle} (fm)$ \\
	\hline
	0 & 289 & $-\infty$ & 0.242 \\
	1.05 & 245 & $-245$ & 0.278 \\
	1.20 & 217 & $-132$ & 0.314 \\
	1.50 & 189 & $-61$ & 0.397 \\
	\end{tabular}
	\caption{\label{bottom_table} Groundstate solutions of the bottomonium at different temperatures for the free energy
${F_1}_T(r)$ . }
\end{table}

\begin{table}[b]
	\begin{tabular}{cccccc}
\hline
$M$ [MeV] & $T/T_c$ &	$l$ & $n$ & $B$ [MeV] & $rms$ [Fm]\\
\hline
1300 & 1.13 & 0 & 0 & -355 & 0.458 \\
	& 1.18 & 0 & 0 & -273 & 0.509 \\
	& 1.27 & 0 & 0 & -121 & 0.618 \\
	& 1.4 & 0 & 0 & -78 & 0.720 \\
	& 1.64 & 0 & 0 & -17 & 1.199 \\
	& 1.95 & 0 & 0 & -1.4 & 3.3606 \\
1752 & 1.13 & 0 & 0 & -462 & 0.389 \\
    & & 0 & 1& - 316  &7.99 \\
	&       & 1 & 0 & -42 & 0.820 \\
	& 1.18 & 0 & 0 & -365 & 0.422 \\
	&	&	0& 1 & -4 & 2.981 \\
	&	&	1 &	0 &	-23 & 1.090 \\
	& 1.27 & 0 & 0 & -195 & 0.477 \\
	& 1.4  & 0 & 0 & -136 & 0.539 \\
	& 1.64 & 0 & 0 & -55 & 0.687 \\
	& 1.95 & 0 & 0 & -22 & 0.936 \\
	\end{tabular}
	\caption{\label{tab:intern_charm}  Charmonium boundstates computed with the internal energy ${U_1}_T(r)$ }
\end{table}

\begin{table}[b]
	\begin{tabular}{cccccc}
\hline
$M$ [MeV] & $T/T_c$ &	$l$ & $n$ & $B$ [MeV] & $rms$ [Fm]\\
\hline
4750 & 1.13 & 0 & 0 & -810 & 0.244 \\
	&	&	& 1 &	-224 & 0.521 \\
	&	&	& 2 &	-5 & 1.752 \\
	&	& 1	& 0 &	-400 &	0.381 \\
	&	& 	& 1	& -31 &	0.924 \\
	&	& 	& 2	& 0 & -118  0.541 \\
	& 1.18 & 0 & 0 & -683 &	0.252 \\
	&	&  & 1 & -168 & 0.596 \\
	&	&  & 2 & -10 & 1.723 \\
	&	& 1 & 0	& -302 & 0.413 \\
	&	&	& 1 &	-30 & 1.068 \\
	&	& 2	& 0	& -68 &	0.659 \\
	& 1.27 & 0 & 0 & -487 & 0.257 \\
	&	&  & 1 & -50 &	0.774 \\
	&	&  1 & 0 & -130 & 0.470 \\
	&  1.4 & 0 & 0 & -385 & 0.274 \\
	&	& &	1 &	-28 & 0.903 \\
	&	& 1 & 0 & -79 & 0.543 \\
	& 1.64 & 0 & 0 & -271 & 0.285 \\
	&	& 1 &  0 & -5 & 0.959 \\
	& 1.95 & 0 & 0 & -212 & 0.295 \\
	& 2.61 & 0 & 0 & -119 & 0.327 \\
	\end{tabular}
	\caption{\label{tab:intern_bottom}  Bottomonium boundstates computed with the internal energy ${U_1}_T(r)$ }
\end{table}

\begin{table}[b]
	\begin{tabular}{cccccc}
\hline
$M$ [MeV] & $T/T_c$ &	$l$ & $n$ & $B$ [MeV] & $rms$ [Fm]\\
\hline	
5100 &	1.13 &	0 &	0 &	-834 &	0.236 \\
	&	&	0 &	1 &	-251 & 0.497 \\
	&	&	  & 2 &	-11 & 1.409 \\
	&	& 1 & 0 & -426 & 0.369 \\
	&	&   & 1 & -47 & 0.822 \\
	&	&   2	& 0 & -145 & 0.514 \\
	& 1.18 & 0 & 0 & -706 &	0.243 \\
	&	&  &	1 &	-188 &	0.564 \\
	&	&  &	2 &	-16	& 1.474 \\
	&	&  1	& 0 & -324 & 0.369 \\
	&	&	& 1 & -40 & 0.963 \\
	&	&   2 &	0 &	-87 & 0.613 \\
	& 1.27 & 0 & 0 & -508 & 0.248 \\
	&	&	& 1  &	-62 & 0.711 \\
	&	&   1 &	0 &	-149 &	0.445 \\
	& 1.4 &	0 &	0 &	-440 &	0.264 \\
	&	&   & 1 &  -37 & 0.818 \\
	&	& 1 & 0	& -94 & 0.509 \\
	&   1.64 & 0 & 0 & -290 & 0.272 \\
	&   &	1 &	0 &	-13 & 0.735 \\
	& 1.95 &	0 &	0 &	-230 &	0.280 \\
	& 2.61 &	0 &	0 &	-135 &	0.306 \\
\end{tabular}
	\caption{\label{tab:intern_bottomII}   Bottomonium boundstates computed with the internal energy ${U_1}_T(r)$ (continued)}
\end{table}

Notice that a Coulomb potential is sufficient to bind an infinite number
of charmonia, and therefore the loss of the linear confinement is not 
sufficient to melt the charmonia states.
A detailed fit of the finite temperature potentials is needed to determine
the melting of charmonia.

%%%%%%%%%%%%%%%%%%%%%%%%%%%%%%%%%%%%%%%%%%%%%%%%%%%%%%%%%%%%%%%%%%%%%%%
%%                                                                   
%%
%%                                   SSS                             
%%
%%                                  S   S                            
%%
%%                                   S                               
%%
%%                                    S                              
%%
%%                                     S                             
%%
%%                                  S   S                            
%%
%%                                   SSS                             
%%
%%                                                                   
%%
%%%%%%%%%%%%%%%%%%%%%%%%%%%%%%%%%%%%%%%%%%%%%%%%%%%%%%%%%%%%%%%%%%%%%%%
\section{The finite temperature static quark potentials}

We assume that the static quark-antiquark potential $V$ is due to the length
$r$ of the flux tube, related the volume $V_{vol}$ of excited QCD vacuum confined into the flux tube.
Our first step consists in assessing $V$, the static quark-antiquark potential,
\begin{equation}
		d V   = - \sigma \, d r  \ ,
\end{equation}
where here  $\sigma $ is a force, generalizing the string tension.
In lattice QCD, with the Polyakov Loop, one computes $F_1$, the Free Energy
\begin{equation}
d  F_1 = - \sigma \, d  r -  S \, d  T   	\ ,
\end{equation}
which equals the potential for isothermal transformations.
To compute $U_1$, the internal energy, one needs to compute the 
entropy $S$ with the total action as well
\begin{eqnarray}
d U_1 &=& - \sigma \, d  r + T \, d  S	\ ,
\nonumber \\
&=& d  F_1- d (TS)\ ,
\end{eqnarray}
which equals the potential for adiabatic transformations.
Naturally close to the phase transition temperature $T_c$, the
transformations are nearly isothermal and, and the potential $V$ is close to
the internal energy $U_1$, while far from $T_c$ the transformations
are nearly adiabatic potential is close to the free energy $F_1$. The interpolating
relation for $V$ between $F_1$ and $U_1$ has been derived by Wong.

Lattice QCD provides potentials and energies, both with quenched (pure gauge) and dynamical (with fermions) for quarks in 
confined system such as mesons, diquarks, baryons, tetraquarks, pentaquarks and hybrids.
Lattice also computes spectral distribution functions, to access directly, say, the bottomonium masses.
However the limitations of the lattice QCD potentials are,
\\
- static potentials only have been computed, 
\\
- a constant shift of the potential is undetermined
\\
- few spin dependent potentials are so far computed,
\\
- lattices have a relatively small volume.
\\
Nevertheless we can extract the potentials from lattice data and use them to compute
the spectra with the Schr\"odinger equation.

%%%%%%%%%%%%%%%%%%%%%%%%%%%%%%%%%%%%%%%%%%%%%%%%%%%%%%%%%%%%%%%%%%%%%%%
%%                                                                   
%%
%%                                   SSS                             
%%
%%                                  S   S                            
%%
%%                                   S                               
%%
%%                                    S                              
%%
%%                                     S                             
%%
%%                                  S   S                            
%%
%%                                   SSS                             
%%
%%                                                                   
%%
%%%%%%%%%%%%%%%%%%%%%%%%%%%%%%%%%%%%%%%%%%%%%%%%%%%%%%%%%%%%%%%%%%%%%%%
\section{Fitting between the T=0 envelope and the finite T saturation}

In the plots for the free energy ${F_1}_T$ and the internal energy ${U_1}_T$ for the different temperatures $T$ as in
Figs. \ref{U1Kacz} and  \ref{F1Kacz}, it is clear that each set of energies is bound from above by a common enveloping function. 

The enveloping functions are respectively the $T=0$ energies, ${U_1}_0$ and  ${F_1}_0$.
Notice that the  $T=0$ free (internal) energy also coincides with the
small distance part of the free (internal) energy functions. 

In what concerns the large distance par of the free (internal) energy, for $T<T_c$ the string tension decreases, while for $T>T_c$ 
the string tension vanishes and the free (internal) energy saturates.

Thus we fit the free (internal) energies in three steps:
\\
- fist we fit the enveloping function
\\
- then we fit the long distance saturation
\\
- finally we match the long distance part to the short distance par of the energy.

We fit the enveloping function $V(r)$ of the free energies ${F_1}_T(r)$ with a constant shift $c$, two Coulomb potentials,
one screened for the short distance and another for the long distance part of the
potential, and a linear potential,
\begin{equation}
V(r) =  c + { - \alpha + \alpha' e^{-\Lambda r} \over r} + \sigma r  \ .
\end{equation}
The parameters are shown in Table \ref{free energy parameters}.

We fit the long distance part of the free energies  ${F_1}_T(r)$  with a saturation function
$M_T(r)$, including a constant shift $d$ and an exponential decay,
\begin{equation}
M_T(r) = d_T - A e^{-\lambda_T \, r } \ .
\end{equation}
The parameters are shown in Table \ref{free energy parameters}.

The interpolation between the short distance $F_0(r)$ and the
long distance $M_T(r)$ is performed with the coupled channel method,
where we choose the lowest eigenvalue of the matrix
\begin{equation}
\left(
	\begin{array}{cc}
	V(r) & w_T \\
	{w_T}^* & M_T(r) \\
		\end{array}
\right) \ ,
\end{equation}
where the interpolating parameter $w_T$ is shown in Table \ref{free energy parameters}.

Finally the free energy is,
\begin{equation}
F_T(r)={ V(r) +M_T(r) - \sqrt{\left[V(r) - M_T(r)\right]^2 + |w|^2 } \over 2} \ .
\end{equation}
The fit is achieved with  5 constant parameters for $V(r)$, and 4 parameters per temperature 
for the long distance part $M_T(r)$ and for the matching parameter $w$.

In what concerns the internal energy  ${U_1}_T(r)$, a fit with similar functions but with
different parameter is performed, except for the higher temperatures that need an extra 
pair of parameters in the saturation function,
\begin{equation}
M_T(r) = d_T - A e^{-\lambda_T \, r }- A' \,  {e^{-\lambda'_T \, r }\over r }\ .
\end{equation}
The parameters of the internal energy are listed in Table  \ref{internal energy parameters}.

%%%%%%%%%%%%%%%%%%%%%%%%%%%%%%%%%%%%%%%%%%%%%%%%%%%%%%%%%%%%%%%%%%%%%%%
%%                                                                   
%%
%%                                   SSS                             
%%
%%                                  S   S                            
%%
%%                                   S                               
%%
%%                                    S                              
%%
%%                                     S                             
%%
%%                                  S   S                            
%%
%%                                   SSS                             
%%
%%                                                                   
%%
%%%%%%%%%%%%%%%%%%%%%%%%%%%%%%%%%%%%%%%%%%%%%%%%%%%%%%%%%%%%%%%%%%%%%%%
\section{Solving the Schr\"odinger equation}

To solve the Schr\"odinger equation
\begin{equation}
	- \frac{ \hbar^2 }{ 2 m } \nabla^2 \psi(\mathbf{r}) + V(r) \psi = E \psi(\mathbf{r})
\end{equation}
with a potential depending only in $r$, we can separate the eigenfunctions as
\begin{equation}
	\psi(\mathbf{r}) = \frac{u(r)}{r} Y_{lm}( \theta, \varphi ) \ ,
\end{equation}
and we get the following equation for the radial component $u(r)$ ( $\hbar = 1$ )
\begin{equation}
- \frac{1}{2 m} \frac{d^2 u_{nl}}{d r^2} + \frac{l(l+1)}{2 m r^2} 
 u_{nl} + V(r) u_{nl} = E_{nl} u_{nl} \ .
\end{equation}
To solve the radial equation we discretize it, with the finite difference
substitution
\begin{equation}
	\frac{d^2 u}{d r^2} \rightarrow \frac{1}{a^2}( u_{i+1} - 2 u_i + u_{i-1} ) \ .
\end{equation}
We also impose the Dirichlet boundary conditions $u_0 = u_N = 0$, consistent
with the radial equation. 
So we get the linear system
\begin{equation}
	H_{ij} u_j = E u_i
\end{equation}
where $H_{ij}$ is tridiagonal.
We solve the linear system for the lowest eigenvalues, by using the inverse iteration method.

The results depend on the mass of the heavy quarks, affecting the kinetic energy and on the
temperature, affecting  the potential. The charm mass and the bottom mas, in different potential 
models, ranges respectively from 1300 MeV to 1752 MeV and from 4750 to 5100 MeV.

Since the lattice potentials have a constant energy shift undefined,
we don't show the total energy of the systems, only the binding energy, 
defined with the difference between the potential at infinity and the 
boundstate energy. For instance for $T=0$, where confinement occurs, 
the binding energy is $-\infty$.

%%%%%%%%%%%%%%%%%%%%%%%%%%%%%%%%%%%%%%%%%%%%%%%%%%%%%%%%%%%%%%%%%%%%%%%
%%                                                                   
%%
%%                                   SSS                             
%%
%%                                  S   S                            
%%
%%                                   S                               
%%
%%                                    S                              
%%
%%                                     S                             
%%
%%                                  S   S                            
%%
%%                                   SSS                             
%%
%%                                                                   
%%
%%%%%%%%%%%%%%%%%%%%%%%%%%%%%%%%%%%%%%%%%%%%%%%%%%%%%%%%%%%%%%%%%%%%%%%
\section{Results}

We now show the results of the boundstate equations.

A fisrt study of the melting of the boundstates can be achieved with
the enveloping potentials, for the internal energy and for the free energy.
Comparing the energy levels of the enveloping potentials with the saturation 
energies at the different temperatures, we estimate the melting temperatures.
Essentially, we can estimate melting to occur when the energy level is close
to the saturation energy.

Thus we first solve the Schr\"odinger equation with the $V(r)$ enveloping 
potential for the charmonium and for the bottomonium and compare the energy 
levels with the potentials obtained in finite temperature. This is depicted 
in Figs. \ref{freeenvelope} and \ref{internalenvelope}.
It occurs that only few of the tempetarures $T>T_c$ will provide binding for the
charmonia, while the bottomonia survives up to higher temperatures. It also
appears that the internal energy may provide binding up to higher temperatures 
than the free energy.

We then study binding for the different potentials extacted fom lattice 
QCD simulations of Kaczmarek
\cite{Doring:2007uh,Hubner:2007qh,Kaczmarek:2005ui,Kaczmarek:2005gi,Kaczmarek:2005zp}.  
with the free energy  (quenched or dynamical) the $J/\psi$ and $\eta_c$ melt at 
$T/T_c\simeq 1.17$ and the groundstate bottomonium melts at $T/T_c\simeq 1.8$.
With the internal energy (quenched) the groundstate charmonium melts at $T/T_c \simeq 1.6$,
while the p-wave excitations, say the $\chi_c$, melt at $T/T_c\simeq 1.1$, while
the bottomonium groundstate melts at $T/T_c\simeq 3.2$, the p-wave 
	$\chi_b$ melts at $T/T_c\simeq 1.7$,
	the d-wave melts at $T/T_c\simeq 1.2$.
Examples of wavefunctions are also show in Figs. \ref{freeenvelope} and \ref{internalenvelope}.
The details of the binding are shown in 
Tables \ref{charm_table} and \ref{bottom_table} for the free energy and in 
Tables \ref{tab:intern_charm}, \ref{tab:intern_bottomII} and \ref{tab:intern_bottomII}
for the internal energy.

%%%%%%%%%%%%%%%%%%%%%%%%%%%%%%%%%%%%%%%%%%%%%%%%%%%%%%%%%%%%%%%%%%%%%%%
%%                                                                   
%%
%%                                   SSS                             
%%
%%                                  S   S                            
%%
%%                                   S                               
%%
%%                                    S                              
%%
%%                                     S                             
%%
%%                                  S   S                            
%%
%%                                   SSS                             
%%
%%                                                                   
%%
%%%%%%%%%%%%%%%%%%%%%%%%%%%%%%%%%%%%%%%%%%%%%%%%%%%%%%%%%%%%%%%%%%%%%%%
\section{Conclusion}

Wong
\cite{Wong:2004zr}
studied how the static potential interpolates between the free energy and
the internal energy at different temperatures. 
Wong showed that at $T\simeq T_c$ the free energy approximates the static
potential, while at $T >> T_c$ the static potential is closer to the internal
energy. 

We study the binding of charmonium and bottomonium, both for the free energy
and for the internal energy, with a finite difference lattice of 200001 points. 
Our study reproduces the charmonium melting temperatures in the literature.

In particular the $J/\psi$ melts above Tc (at 1.6 Tc according to lattice results)  
but the excited charmonium $\chi_c$ or $\psi^*$ are probes that melt just above Tc.
New in our results are the detailed studies of the bottomonium, relevant for
the LHC, with a melting as high as $T/T_c \simeq 3.2$. 

Quantitative Puzzles remain in understanding the the mass shifts $>$ 300 MeV , 
unseen in the experimental data, and in the importance of the
the spin dependent potentials, missing in the lattice QCD data.

Possible future efforts may be to,
\\ - calibrate the best we can the T-dependent quark-antiquark potentials,
\\ - apply the potentials to chiral symmery breaking/restoration,
\\ - apply chiral restoration to the light quark mesons $\rho, \omega, \phi$ and $\pi, K$.

\acknowledgements
We are very grateful to Olaf Kaczmarek for providing
his results for the static free energy and internal energy
in lattice QCD.

This work is supported by Funda\c{c}\~ao para a
Ci\^encia e a Tecnologia under the grants PDCT/FP/63437/2005
and PDCT/FP/63923/2005.

%bbbbbbbbbbbbbbbbbbbbbbbbbbbbbbbbbbbbbbbbbbbbbbbbbbbbbbbbbbbb
%bbbbbbbbbbbbbbbbbbbbbbbbbbbbbbbbbbbbbbbbbbbbbbbbbbbbbbbbbbbb
%bb
%bb
%bbbbbbbbbbbbbbbbbbbbbbbbbbbbbbbbbbbbbbbbbbbbbbbbbbbbbbbbbbbb
%bbbbbbbbbbbbbbbbbbbbbbbbbbbbbbbbbbbbbbbbbbbbbbbbbbbbbbbbbbbb

\end{document}